\documentclass[aps,pre,superscriptaddress,twocolumn,floatfix]{revtex4-1}
\usepackage{graphicx}
\usepackage{dcolumn}
\usepackage{bm}
\usepackage{hyperref}
\usepackage{amssymb}
\usepackage{amsmath}
\usepackage{array}
\usepackage[normalem]{ulem}

\usepackage{rotating} 

\usepackage{siunitx}

\usepackage{natbib}
\usepackage{epstopdf}
\usepackage{color}
\usepackage{makecell}
\usepackage[table]{xcolor}
\usepackage{tabularx}
\usepackage{ragged2e}

\usepackage{lmodern}
\usepackage[latin1]{inputenc} 

\graphicspath{{figure/}} 

\definecolor{lightgray}{gray}{0.9}
\definecolor{darkgray}{gray}{0.60}
\newcolumntype{C}{c<{\kern\tabcolsep}<{\kern\tabcolsep}@{}}
\begin{document}

\newcommand{\be}{\begin{equation}}
\newcommand{\ee}{\end{equation}}
\newcommand{\del}{\partial}

\let\oldAA\AA
\renewcommand{\AA}{\text{\normalfont\oldAA}}

\newcommand{\LL}[1]{\textcolor{red}{{\bf LL:} #1}}
\newcommand{\LH}[1]{\textcolor{blue}{{\bf LH:} #1}}
\newcommand{\CH}[1]{\textcolor{cyan}{{\bf Ch:} #1}}

\preprint{APS/123-QED}

\newcommand{\udem}
{D\'{e}partement de Physique and Regroupement
 Qu\'{e}b\'{e}cois sur les Mat\'{e}riaux de Pointe, 
Universit\'{e} de Montr\'{e}al, C.P. 6128, Succursale Centre-Ville,
 Montr\'{e}al, Qu\'{e}bec, Canada H3C~3J7}
\newcommand{\nrc}
{National Research Council of Canada, Ottawa, On.,  Canada K1A 0R6}

\title{
Equation of state, phonons, and lattice stability of
ultra-fast warm dense matter}

\author{L. Harbour}\affiliation{\udem}
\author{M. W. C. Dharma-wardana}\affiliation{\nrc}
\author{D. D. Klug}\affiliation{\nrc}
\author{L. J. Lewis}\affiliation{\udem}

\date{\today}

\begin{abstract}  
Using the two-temperature  model for  ultrafast
matter (UFM), we compare the equation of state, pair-distribution
functions $g(r)$, and phonons using the neutral pseudoatom (NPA) model with
results from  density-functional theory (DFT) codes and molecular-dynamics (MD)
simulations for Al, Li and Na.  The NPA approach uses state-dependent
first-principles pseudopotentials from an `all-electron' DFT calculation with
finite-$T$ XCF. It provides  pair potentials, structure factors,  the
 `bound' and `free' states, as well as a  mean
ionization $\bar{Z}$ unambiguously. These are not easily accessible {\it via}  DFT+MD
calculations which become prohibitive for $T/T_F$ exceeding $\sim 0.6$, where
$T_F$ is the Fermi temperature.  Hence, both DFT+MD and NPA  methods can be
compared up to $\sim 8$ eV, while higher $T$ can be addressed  ${\it via}$
the NPA. The high-$T_e$ phonon calculations raise  the question of UFM lattice
stability and surface ablation in thin UFM samples.  The ablation forces in a
UFM slab are used to define an ``ablation time'' competing with phonon
formation times  in thin UFM samples. Excellent agreement  for all properties
is found between NPA and standard DFT codes,  even for Li where a strongly
non-local   pseudopotential is used in DFT codes. The need to use
pseudopotentials appropriate to the ionization state $\bar{Z}$  is emphasized.
The effect of finite-$T$ exchange-correlation functional is illustrated via its effect on the pressure and
the electron-density distribution at a nucleus.
\end{abstract}

\maketitle

\section{Introduction.} 
\label{intro-sec}
The equation of state (EOS) of common thermodynamic phases of matter 
is well understood. However, recent 
laser and shock-wave experiments have accessed novel ultrafast regimes
 of density and
temperature which are of great theoretical and technological interest. 
The same physics appears during the injection of
hot carriers in field-effect transistors and other nanostructures.
Topics like inertial-confinement fusion~\cite{bib2},
Coulomb explosions~\cite{bib3}, space re-entry shielding, laser machining
and ablation~\cite{bib1} involve such regimes of warm dense matter
(WDM). However, elementary approaches
cannot be applied since the Coulomb coupling constant
$\Gamma$, i.e., the ratio of the Coulomb energy to the kinetic energy, is
larger than unity.  The electrons may range from degenerate
to Boltzmann-like, with $T/E_F\sim 1$ or larger,
 where $T$ is electron temperature
 in energy units,
while $E_F$ is the Fermi energy. This
causes a prohibitive increase in basis sets that
span the many excited electronic states. 
WDMs pose a theoretical
challenge for rapid accurate computations
of properties like pressure, heat capacity, phonons and  conductance
needed even for equilibrium WDMs.

A class of WDMs known as ultra-fast matter (UFM) is produce when energy is
deposited using an ultrafast  pulsed laser  on a metal surface
\cite{Ping10}. The light couples strongly to the mobile electrons  which 
equilibrate on  femtosecond timescales, to a temperature $T_e$ (as high
as many eV) while the much heavier ions and their strongly-bound core electrons
remain essentially at their initial temperature $T_i$, i.e.,  usually the room
temperature $T_r$. This two-temperature WDM ($2T$-WDM) phase with $T_e>T_i$
remains valid for  timescales $t$  such that  $\tau_{ee} < \tau_{ii} < t <
\tau_{ei}$, where $\tau_{ee}$, $\tau_{ii}$ and $\tau_{ei}$ are the
electron-electron, ion-ion and electron-ion temperature relaxation times,
respectively. It has been shown  for near-solid densities
 that $\tau_{ei}$ is of
the order of picoseconds, and orders of magnitude longer than $\tau_{ee}$ and
$\tau_{ii}$~\cite{Milchberg88,bib7}. For WDMs with $\theta=T/E_F$ small, similar
 relaxation times hold
as seen in calculations for typical systems~\cite{XRTS2016}.
 Experiments using femtosecond  pump-probe
techniques~\cite{bib4,chen2013} provide data for quasi-equilibrium analogues of
free energy and pressure, transport and relaxation processes. While many UFM
samples do not conform to the $2T$ model (e.g., as in Medvadev 
{\it et al.}~\cite{Medvadev2011}), the $2T$ model  provides a great
simplification when it holds. Even for UFMs, theory and
experiment are quite challenging as the system transits rapidly from a solid to
a plasma depending on the pump energy. Hence a theoretical model that
encompasses a wide range of material conditions is needed to describe the time
evolving system as a series of static  2$T$ systems. The `quasi-equilibrium' 
theory is applied to each static picture of the time evolving system.

In this work, we use the neutral pseudoatom (NPA) model, in the form given by 
Perrot and Dharma-wardana \cite{Dagens1, Dagens2, PerrotBe, PDW95, MuriDW2008,
MuriDW2013}, to study the $2T$-WDM regime of a few nominally \textit{simple
metals}, \textit{viz}.,  aluminum, lithium and sodium. These  are ``simple'' at
ambient conditions  since their valence electrons are  ``free-electron like'' and
energetically separated from the core electrons. The number $\bar{Z}$ of valence
electrons per atom (mean ionization) for Al, Li, and Na is 3, 1, and 1,
respectively. Furthermore, if the matter density is $\rho$, each ion can be
assigned a spherical volume with the Wigner-Seitz (WS)  radius
$r_{ws}=(3/4\pi\rho)^{1/3}$, and it can be shown for Al, Li, Na that
 the bound-electron core has a radius $r_c$ such that it is well inside
 the WS sphere for the temperatures studied here (see Sec.~\ref{pseudo.sec}).
 In such cases,
the definition of $\bar{Z}=N-n_b$, where $n_b$ is the number of bound electrons in
the core, is unambiguous, clear and is a physically measurable quantity, e.g., using
X-ray Thomson scattering~\cite{GlenzerRevMod}.
 In the case of equilibrium WDM, the NPA-calculated $\bar{Z}$ for Al and Li  remains 3 and 1
in the range $0<T_e<8$ eV whereas in the case of sodium,  $\bar{Z}$ rises to 1.494 by T=8 eV and 1.786 by 10 eV. The case of Na provides us an example of
a typical variation of $\bar{Z}$ very common in equilibrium WDM systems and handled
without any ambiguity and with thermodynamic consistency by the NPA approach
coupled with determinations of the ion-ion $g(r)$ using the NPA pair-potentials. However, in the case of UFM which is the scope of this work, $\bar{Z}$ is kept unchanged for all three elements through the $0<T_e<8$ eV temperature range.

The NPA model replaces the interacting \textit{many-nuclear} and 
{\it many-electron} problem by an effective non-interacting 
\textit{single-nuclear} and {\it single-electron} formulation  where the
many-body problem is reduced using  finite-$T$ density functional theory 
(DFT)~\cite{bib8,bib9}. The NPA charge  densities are used to construct $2T$
pseudopotentials  and  effective ion-ion pair potentials. The method takes into
account particle correlations at the pair-density level and beyond using
density-functional methods via exchange-correlation functionals for electrons,
and ion-correlation functionals for ions in a decoupled step which uses a
classical integral equation or molecular dynamics. The NPA framework is well
adapted to treating metallic systems ranging from solids to liquids or plasmas
at very high or low compressions, and from $T$=0 to several keV. The importance
and relevance of the NPA lies in its accuracy, flexibility, and  computational
rapidity compared to DFT coupled to molecular dynamics (MD) methods (DFT+MD). 
However, the NPA, as used here, is  inapplicable  when inner-shell electrons
(e.g., $d$-electrons) play a role in the ion-ion interactions (e.g., as in
transition metals). A simple metal becomes `complex' when its
electronic bound states extends beyond its WS radius $r_{ws}$. This is not
a short-coming  but a strength of the model which signals the need
for multi-ion contributions into the theory in such ranges of
temperature and pressure. In such regimes, discontinuities in $\bar{Z}$ where
some are spurious  may appear
unless suitable electron-ion XC-correlation potentials are included in
the theory~\cite{Furutani90}. Furthermore {\it transient} molecule
 formation can be successfully handled~\cite{carbonCDW16} within the
 NPA as it allows for binary ion-ion correlations.

We compare our $2T$-NPA predictions with those from solid-state DFT
electronic-structure codes   such as ABINIT~\cite{bib19} and VASP~\cite{VASP},
which use MD to evolve the finite-$T$ ionic structures. These  codes are
primarily designed for $T_i=T_e=0$ situations, and solve the
\textit{multi-nuclear} Kohn-Sham equations in a plane-wave basis, using $T=0$
pseudopotentials to reduce the number of electrons needed in the simulations.
The solid, liquid or plasma is treated as a {\it periodic solid} in a
simulation box (``supercell'') containing $N$ nuclei, with $N$ being $\sim$100.
A finite $T_e$  Fermi-Dirac distribution for electron occupation numbers is
used, along with $T=0$ pseudopotentials and $T=0$ exchange-correlation
functionals (XCF). The number of electronic bands required to access high $T_e$
increases rapidly with $T_e$ and becomes prohibitive for $T_e/E_F$ greater than
$\sim 1$. This method generates energy bands for the periodic solid where as in
reality there are {\it no such band structure} in  liquids and plasmas. This 
artifact is overcome by
generating electronic-structure calculations for  many static ionic
configurations via MD simulations and averaging over a large number of them.

DFT+MD provides only a ``mean ionization'' for the whole $N$-ion
supercell; it cannot provide, e.g.,  the composition of an equilibrium
mixture of specific charge states of ions in a C, H ``plastic'' at, say, 1
eV.   Furthermore, VASP and ABINIT currently only implement  the zero-$T$ XCF 
even though finite-$T$ parametrizations have been available for some time,
e.g.,  the evaluation of finite-$T$ bubble diagrams~\cite{bubble1, bubble2},
from  the work of Iyetomi and Ichimaru~\cite{Iye}, Perrot and Dharma-wardana
(PDW)~\cite{PDWfxc} and from Feynman-path methods by Brown \textit{et
al}~\cite{Brown} parametrized recently by Karasiev \textit{et
al}~\cite{Karasiev}. The present NPA calculations  are done with the PDW
finite-$T$ XCF which is in close agreement with the quantum simulations of
Brown {\it et al.}~\cite{cdw-cpp15}. In most cases finite-$T$ XC effects
 contribute only small corrections and DFT+MD provides valuable benchmarks
 for testing other methods.

The NPA method is summarized in section~\ref{sec-npa} where we emphasize its
application to the $2T$ regime.  Resulting $2T$ pair potentials ($2T$PP),
quasi-equilibrium phonon dispersions and pair distribution functions (PDF)
$g(r)$ are presented in Sec.~\ref{results.sec}. The phonon calculations confirm
the  results and also validate the meV accuracy of the NPA method. The NPA
$g(r)$ calculations for normal and compressed Li ($\sim$ up to a compression of
2) show that the {\it local} pseudopotential for Li is successful. Here we
compare the ion-ion structure factor $S(k)$ with the  simulations of Kietzmann
{\it et al}. Having confirmed the accuracy of the pseudopotentials and pair
potentials, the  2$T$-thermodynamic properties, such as the quasi-pressure, are
also presented. These are compared with the values for  systems in thermal
equilibrium. Discussions about phonon formation times in $2T$ systems, the
role  of finite-$T$ XC-contributions in the $2T$-EOS calculation, and the
choice of  suitable pseudopotentials in \textit{ab initio} finite-$T$
simulations are also presented.   

\section{The neutral pseudoatom model.}
\label{sec-npa}
\subsection{General description of the model.}  Several average-atom models and
NPA models have been proposed, even in the early literature \cite{Ziman}. 
Many of these are intuitive cell models  and are not true
DFT models. A rigorous DFT formulation of a NPA model at $T=0$ was first  used
for solids by Dagens \cite{Dagens1, Dagens2}. There the treatment of the ion
distribution was developed in the traditional manner as providing a fixed
external potential; Dagens showed that the NPA results at $T=0$ agree closely
with the band-structure codes available at the time.  A finite-$T$ version was
given in several papers by Perrot~\cite{PerrotBe} and
Dharma-wardana~\cite{DWP1, PDW2, PDW3}.  In Ref.~\cite{DWP1}, the ion
distribution $\rho(r)$ itself was treated within DFT using the property that
the free energy $F[n,\rho]$ is a functional of {\it both} $n(r)$ and $\rho(r)$
simultaneously. A classical DFT equation for the ions  and an ion-correlation
functional, $F_c^{ii}(\rho)$, approximated as a sum of hypernetted-chain (HNC)
diagrams plus bridge diagrams, was introduced, {\it without} invoking a
Born-Oppenheimer approximation or treating the ions as providing a fixed
external potential~\cite{IlCiacco93}. Exchange-correlation functionals 
$F_{xc}^{ei}(\rho)$ for
electron-ion interactions were also introduced  although neglibible
in common materials. This puts the NPA approach on a very
rigorous DFT footing where approximations enter in modeling the ion-correlation
functional, just as in the case of the electron DFT problem for the electronic
XCF.

However, in the following  we present the theory in terms of the more familiar
superposition picture. We consider a system of ions located at sites
$\mathbf{R}_i$ at temperature $T_i$ and average density $\rho$, interacting
with a system of electrons at temperature $T_e$ and average density $n$.  The
\textit{multi-center} problem is reduced to a simplified \textit{single-center}
problem where the total electron density $n(r)$ is  regarded as the
superposition of single-site densities such that $n(r)=\sum_i
n_i(r-\mathbf{R}_i)$.  In contrast to ion-sphere (IS) models like those used in
Purgatorio~\cite{Purgatorio}, or Piron and Blenski~\cite{Blenski2011}, Starrett
and Saumon~\cite{Starrett2013}, the single-site free-electron density $n_f(r)$
extends over the whole of space, approximated by a correlation
sphere~\cite{DWP1} of radius $R_c$ which is of the order of 10 ionic 
Wigner-Seitz radii. All particle correlations are assumed to have died out when
$r\to R_c$. This $R_c$ is similar to the linear dimension of the simulation box
of  a DFT+MD simulation which has to be as big as possible. However, in
practice the charge distribution used in DFT+MD simulations spreads over a
volume of about  100 ions.  In contrast, the NPA  correlation sphere with 
$R_c \simeq
10r_{ws}$  extends over $\{R_c/r_{ws}\}^{3}$, i.e., the volume covered by
$\sim$1000 ions. The calculation of course uses only one nucleus, but its
charge  density overlaps the space of  some 1000 atoms, and this is
crucial to getting the right pair-potentials with long-range Friedel
oscillations, and to satisfy the Friedel sum rule~\cite{DWP1}. The IS-models
cannot satisfy the Friedel sum rule. At higher temperatures where particle
correlations are weak, $r_c$ may be reduced to, e.g.,  $5r_{ws}$, but the
results are independent of $R_c$, and $R_c$ is {\it not} an optimization
parameter. 

The ion distribution $\rho(r)=\rho g_{ii}(r)$
contains the full ion-ion PDF, $g(r)$, when seen from any site taken as the
origin. It is found that in most cases it is sufficient, as far as the
 bound-electron  structure is concerned, to approximate $g(r)$
by a spherical cavity $c(r)$ of radius $r_{ws}$ and total charge $\bar{Z}$
centered on the ion, followed by a uniform positive density $\rho$  for
$r>r_{ws}$. As mentioned below, unlike in IS-models, its effect will be
 subtracted out (as a ``cavity correction'') to obtain the response of a
 uniform electron gas to the nucleus.
Thus have: 
\be
\label{cavity.eq}
c(r) = n[H(0)-H(r-r_{ws})], 
\ee
where $H(r)$ is the Heaviside step function.
Initially $\bar{Z}$ is unknown but its value is obtained self-consistently from
the iterative Kohn-Sham procedure. The single-site electron density is written as 
$n_i = \Delta n_i + m_i$ where  $m_i$ is the cavity correction
  and $\Delta n_i$ is the electron pile-up obtained by
the DFT calculation for the electrons in the external
 potential $V_\text{ext}$ given by
\be
V_{\text{ext}}(r) = -\frac{Z}{r}+ \frac{1}{|\mathbf{r}-\mathbf{r}'|}
\star c(\mathbf{r}')
\ee
where the symbol $\star$ means integration over all space. Here $Z=Z_n$ is the
nuclear charge. The positive
background with the WS-cavity, the nucleus at its center and the free-electron charge density filling the whole correlation sphere constitute the
neutral pseudoatom~\cite{PDW2,PDW3}. The WDM system is made up of superpositions
of such neutral-pseudo atoms correlated to give the ion-ion $g(r)$, with the cavity contributions subtracted out.

For simple metallic systems, this cavity model that defines the extent of the
bound states is sufficient to produce
physically accurate results and is mathematically convenient, as shown in the
papers by Dagens or those of Perrot and Dharma-wardana cited above. Thus, to
compute the cavity correction $m(r)$, we assume that the electrons respond
linearly to the cavity $c(r)$, {\it viz.}, in Fourier space,
\be
m(q) = -V(q)c(q)\chi_{ee}(q, n, T_e).
\ee
Here, $V(q)=4\pi/q^2$ is the Coulomb potential and $\chi_{ee}$ is the
interacting-electron response function at the electron density $n$
 and temperature $T_e$. To go beyond the random phase approximation (RPA),
 we use the following finite-$T$ response function:  
\be
\label{response.eq}
 \chi_{ee}(q, n, T_e)=\frac{\chi_0(q,n,T_e)}{1-V(q)[1-G(q)]\chi_0(q,n,T_e)},
\ee
with $\chi_0$ the finite-$T$ Lindhard function and
$G(q)=G(q,T_e)$ a local-field correction (LFC) defined as:
 \be
G(q) = \left(1-\frac{\gamma_0}{\gamma}\right)\left(\frac{q}{k_{\text{TF}}},
\right)^2.
 \ee
In the above, the Thomas-Fermi wave vector  $k_{\text{TF}}= \sqrt{6\pi n/E_F}$,
is defined by the Fermi energy of the system $E_F=1/(\alpha r_s)$  where $r_s$
is the electron WS radius and  $\alpha=(4/9\pi)^{1/3}$. The
finite-$T$ interacting  electron compressibility $1/\gamma=
n^2\del^2[nf(r_s,T_e)]/\del n^2$  is determined from the homogenous electron
gas free energy per electron  $f(r_s,T_e)$,  as given in Eq.\ref{free_elec},
which include a finite-$T$ XC contribution $f_{xc}$. The non-interacting
electron compressibility $\gamma_0$ is obtained by setting $f_{xc}=0$.

The simplicity of the NPA model rests on decomposing the total charge
distribution into a superposition of single-center distributions. If the
ion-ion structure factor $S_{ii}(q)$ is known, any total electron charge
distribution $n_t(q)$ can always be written as a convolution of the
$S_{ii}(q)$ with some effective single-center charge distribution $n(q)$, even
for transition metals or systems with resonant levels; but partitioning the
electron contributions from states that extend beyond their WS cells without
correctly including the physical interactions is not sufficient. Furthermore, 
a `simple metal' at one temperature may behave as a `transition-metal' at
another temperature when a shell of electrons begins to transit to the
continuum, and {\it vice versa}. If the system is of such low density that
$r_{ws}$ is larger than the bond length of a possible dimer (e.g., Li$_2$),
then the dimer itself will be contained within the WS sphere, and in such
cases the NPA model fails; a more elaborate ``neutral-pseudomolecule''
approach or the use of suitable electron-ion XC-potentials 
$F_{xc}^{ei}(n,\rho)$ is then needed. We
do not examine such non-simple WDMs in this study. Similarly, at high
densities, WDM-Li shows complex phases containing  persistent Li$_4$
clusters~\cite{bonev2008}, and the simple NPA model needs modifications. In
the present case, a single-center decomposition is physically transparent if
the bound electron core is unambiguously confined within the WS sphere of the
ion. We discuss in the results section (sec.~\ref{pseudo.sec}) the variation
of the $\bar{Z}$ of Na which changes from unity at low $T$ to 1.49 by $T=8$
eV. The occupation number in the 2$p$ level begins to decrease, while its
radius slightly decreases, and hence there is no ambiguity in estimating
$\bar{Z}=Z-n_b$ where $n_b$ are all the bound electrons compactly contained
well inside the WS-sphere.   That is, the electron density pileup $\Delta n_i$
can be clearly divided into bound and free parts such that  $\Delta n_i = n_b
+ n_f$.  Once this division is achieved the interaction of an electron with
the nucleus plus its core can in most cases be replaced by a pseudopotential
$U_{ei}$ which is a weak scatterer because it is constructed using linear
response; this is given by:  
\be
\label{eq-pseudopot}
U_{ei}(q)=n_f(q)/\chi_{ee}(q,r_s,T_e),
\ee
where $\chi_{ee}$ is provided by Eq.~\ref{response.eq}. 

Even though  linear response is used, the resulting pseudopotential includes 
non-linear effects since $n_f(q)$ is the fully non-linear free-electron density
obtained from  DFT.  Only a range of $q$ between zero to slightly above $2k_F$ 
(depending on $T_e$) needs to be included as the large-$q$ behavior
(short-range in $r$, i.e., inside the core) is not relevant.  The resulting
pseudopotential is valid only if it satisfies the relation
$U_{ei}/(-\bar{Z}V(q)) \leq 1$. Unlike the pseudopotentials used in VASP,
ABINIT and similar DFT codes, this  linear-response pseudopotential does not
require solving a Schr\"{o}dinger equation.  It is a state-dependent {\it
local} pseudopotential that can be fitted to, say a Heine-Abarankov form for
convenience (see Shaw and Harrison \cite{HA}).  This has a constant core
potential $V_{\text{HA}}= D$ for $r < r_c$ and it is Coulomb-like,
$V_{\text{HA}}= -\bar{Z}/r$ for $r > r_c$. However, such a fitting is not
needed except to conveniently report the pseudopotential and to quantify the
core radius associated with the potential. In our NPA calculations we use
the numerical form of $U_{ei}(q)$ directly. 

The  pseudopotential calculated at $T_i$ can be used to form a $2T$ ion-ion
pair potential ($2T$PP) with ions at $T_i$ and electrons at $T_e$, since it is
a sum of the direct Coulomb interaction and the indirect interaction via the
displaced-electron charge, {\it viz.}\
\be
\label{eq-pairpot}
U_{ii}(q,T_{i},T_e) = -\bar{Z}^2(T_i) V(q) + |U_{ei}(q)|^2\chi_{ee}(q, T_e).
\ee
This procedure is valid because $\bar{Z}$ remains unchanged in  UFM  since
the bound core of electrons remains at the initial ion temperature for times
$t<\tau_{ei}$. If $T_e$ is large enough to change $\bar{Z}$, be it for UFM
or equilibrium systems, then the pseudopotential has to be re-calculated using
an NPA calculation at the needed temperature.
  
At low $T_e$, the Friedel oscillations in the electron density resulting from
the sharp discontinuity at $k=2k_F$ in $\chi_{ee}(q)$ produce oscillations in
the pair potential $U_{ii}(r)$. These lead to  multiple minima in the ion-ion
energy which contribute to the maxima in $g(r)$.  Such physically  important
features are not  found in ``Slater-sum'' approaches~\cite{Slatersum}
 to finite-$T$ potentials,
in `Yukawa-screening' models~\cite{YukawaScr, XRTS2016}, or in
 Gordon-Kim models~\cite{Gordon-Kim72}. Furthermore, the charge densities
 restricted to the WS-sphere used in IS-models cannot capture such 
long-range effects.
 Our  NPA pair potential can
be used to study phonons in the system or to generate the ion-ion $g_{ii}(r)$
and corresponding structure factor $S_{ii}(k)$ when necessary. The ion
subsystem in a UFM is clamped at $T_i\sim 300$K when Al, Li, and Na are
crystalline metals. Hence the ion-ion pair distribution function is simply
given by the relation
\be
g_{ii}(\mathbf{r}) = \frac{1}{4\pi \rho}\sum_{\{  i \} }
\delta(\mathbf{r}-\mathbf{R}_i).
\ee
The summation is over the crystal lattice, permitting a simple computation of
the ion contribution to the quasi free energy and pressure from the $2T$ pair
potential.
\subsection{The NPA quasi thermodynamic relations.}
\label{sec-thermo}
The total free energy $F$ of the $2T$ system given by the NPA is
\begin{equation}
F = F_{\text{emb}}  +F_{\text{cav}}+ F_{\text{heg}} + F_{\text{ion}},
\end{equation}
where $F_{\text{heg}}$, $F_{\text{emb}}$, $F_{\text{cav}}$, and 
$F_{\text{ion}}$ are respectively the free energy contribution of the
interacting homogeneous electron gas (HEG), the embedding free-energy of the
NPA into the electron gas, the correction from the cavity, and the ion-ion free
energy. The only parameters of this model are the nuclear charge $Z$, electron temperature $T_e$
and the HEG density $n$ such that the average ion density $\rho=n/\bar{Z}$,
itself determined by the ion temperature $T_i$. We discuss these four terms
below, using Hartrees with $\hbar=m_e=|e|=1$.

(i) The embedding energy $F_\text{emb}$ is the difference between the free energy
of the electron gas containing the central ion and the unperturbed HEG; thus
\begin{align}
\nonumber &F_\text{emb} = T[n+\Delta n(r)] -
T[n]-\int\frac{\bar{Z}}{|\mathbf{r}|}\cdot 
[\Delta n(\mathbf{r}) + c(\mathbf{r}))]d\mathbf{r}\\
& +\frac{1}{2}\int\frac{[\Delta n(\mathbf{r})+c(\mathbf{r})]}{|\mathbf{r}-\mathbf{r}'|}\cdot[\Delta n(\mathbf{r}')+c(\mathbf{r}')]d\mathbf{r}d\mathbf{r}',
\end{align}
with $T[n]$ is the electron kinetic energy.

(ii) The cavity correction $F_\text{cav}$ is computed from the total screened
Coulomb potential $V(r)$ resulting from the total electron displacement $\Delta
n(r)$:
\be
V^*_i(\mathbf{r}) = \int\frac{[c(\mathbf{r}') + \Delta n(\mathbf{r}') -\bar{Z}\delta(\mathbf{r}'-R_i)]}{|\mathbf{r}-\mathbf{r}'|}d\mathbf{r}'.
\ee 
Since each cavity involves a charge deficit $\eta(r)=n-c(r)$, the cavity
correction is  
\begin{align}
F_\text{cav} =& - \frac{1}{2}\int\frac{\eta(\mathbf{r})\cdot[c(\mathbf{r}')-m(\mathbf{r}')]}{|\mathbf{r}-\mathbf{r}'|} d\mathbf{r}d\mathbf{r}' \\
&\nonumber + \int \eta(\mathbf{r})\cdot V^*(\mathbf{r})d\mathbf{r}.
\end{align}

(iii) The free energy of the HEG $F_\text{heg}$ is written as
\be
\label{free_elec}
F_\text{heg} =\bar{Z}f(n,T_e)= \bar{Z}[f_0(n,T_e)+
f_{xc}(n,T_e)],
\ee
where $f_0$ and $f_{xc}$ are respectively the non-interacting and
exchange-correlation free energies per electron at the density $n$ and
temperature $T_e$. To compute $f_0$, we use the thermodynamic relation
$f_0=\Omega_0/nV+\mu_0$, where $\Omega_0$ and $\mu_0$ are the
non-interacting  grand potential and the chemical potential, respectively.
 
We emphasize that the NPA-Correlation-sphere model uses the non-interacting
$\mu_0$ associated with the mean electron density $n$ as required by DFT theory.
In IS models the known matter density defines the
Wigner-Seitz cell, and the free electrons are confined in it, and the corresponding $\mu$ is determined by an integration within the WS-sphere
(e.g., see Eq. 1 of Faussurier~\cite{Faussurier2014}), leading to a value of $\mu\ne\mu_0$.
In contrast,  the mean electron density $n$, the nuclear charge $Z_n$ and the temperature $T$ are the only inputs to the NPA code.
The computation outputs the corresponding mean ion density $\rho$ and 
$\bar{Z}=n/\rho$.  A series of calculations are done in a range of
$n$ and the specific $n$ which gives the physical ion density, 
viz.,  $\rho$ is selected.
For a given electron density $n$ and temperature $T_e$, the non-interacting
 chemical potential $\mu_0$ is obtained by satisfying the relation
\be
n=(\sqrt{2}/\pi^2)T_e^{3/2}\ I_{1/2}(\mu_0/T_e),
\ee
while, using this $\mu_0$, the non-interacting part of the grand potential
 is given by
\be
\label{eq-om0}
\Omega_0/V  = (2\sqrt{2}/3\pi^2)T_e^{5/2}\ I_{3/2}(\mu_0/T_e),
\ee
with $I_\nu(z)$ the Fermi-Dirac integral of order $\nu$. Note that only the
non-interacting chemical potential, viz., $\mu_0$ appears in the DFT-level
occupations of the NPA model since DFT theory maps the interacting electrons to
a system of non-interacting electrons at the {\it interacting density} (see
also Ref.~\cite{DWP1}). 

The XC contribution $f_{xc}$ is computed directly from the PDW parametrization
at the given $r_s$ and $T_e$. The total free energy per electron of the
interacting HEG is the sum of $f_0$ and $f_{xc}$.

(iv) The ion-ion interaction energy is given explicitly by the pairwise
summation over the pair potential $U_{ii}$ as defined at Eq.(\ref{eq-pairpot}):
\be
F_\text{ion}= \frac{1}{V}\frac{1}{2}\sum_{\{ i\neq j\}} U_{ii}(|\mathbf{R}_i-\mathbf{R}_j|),
\ee
where the sum is over the positions of the ions in their initial crystal
configuration. This is the only term in $F$ that depends explicitly on the ion
structure.

Both the cavity correction and the embedding energy involve the ion with its
bound core of electrons held at the temperature $T_i$, while the electrons are
at $T_e$. The numerical results are insensitive to using a simple NPA
calculation with even the core at $T_e$, if the the bound-state
 occupancies (and thus $\bar{Z}$) remain virtually unchanged.

The quasi-equilibrium pressure of the system is obtained by the appropriate
density derivative of the ion-structure independent free energy terms while
 the structure-dependent ion-ion contribution is given by the viral equation
\begin{align}
P&=n^2 \frac{\del}{\del n}(F_\text{heg}+F_\text{emb}+F_\text{cav}) \\
\nonumber &-\int g_{ii}(\mathbf{r})\left( \frac{3}{r}\frac{\del}{\del r}-n^2 
\frac{\del}{\del n}\right)U_{ii}(\mathbf{r})d\mathbf{r}.
\end{align}
The explicit electron-density dependence of the ion-ion
pair potential is taken into account in computing the pressure~\cite{Hansen}. 
Analytical
results can be obtained for the terms
\begin{align}
&P_\text{emb} = -\int \eta(\mathbf{r})\cdot V^*(\mathbf{r})d\mathbf{r} \\
&P_\text{cav} = -\bar{Z}V^*(r_{ws})
\end{align}
whereas other derivatives have to be done numerically.

\section{Results.}
\label{results.sec}
We used the NPA model to determine the properties of $2T$-WDM as produced by
femtosecond laser pulses interacting with three common metals in their usual
solid state, viz., aluminum, lithium and  sodium, with electron densities such
that $r_s$ is 2.07, 3.25,  and 3.93 a.u., corresponding to $\bar{Z}$ = 3, 1 and
1, respectively. Note that the $\bar{Z}$ for Na deviates from unity for $T>3$ eV.
The ion density is kept constant in the calculations for isochoric sodium.
We present the $2T$ ion-ion pair-potentials, non-equilibrium
phonon dispersion curves and pressures for varying $T_e$, while the ions remain
cold at $T_i=$0.026 eV (300K).

\subsection{Ion-ion pair potentials.}
The first step within our UFM model is to compute the equilibrium (at room
temperature, $T_e=T_i=0.026$ eV) free-electron density $n_f(q)$ from the NPA
calculation. The pseudopotential $U_{ei}(q)$ at $T_e=T_i$ can then be obtained
using Eq.~\ref{eq-pseudopot}. This pseudopotential is an atomic property that
depends on $\bar{Z}$ and on the core radius given the ionic $r_{ws}$, which 
is then used to construct ion-ion pair potentials $U_{ii}(q,T_e)$ at any $T_e$
via Eq.~(\ref{eq-pairpot}). For this the electron response at  $T_e \ne T_i$
is used. This method is simpler and numerically almost indistinguishable from
calculating the pseduopotential from a full 2$T$-NPA procedure where the core
electrons are held frozen at $T_i$ and $n_f(q,T_e)$ is calculated from the
Kohn-Sham equation, with $\bar{Z}$ remaining unchanged. The agreement between
the two different ways of calculating the 2$T$ potentials provides a strong
check on our calculations. Furthermore, while pair potentials  cannot be
easily extracted from {\it ab initio} calculations,  the NPA model provide
this physically important quantity. 

\begin{figure}[h]
\includegraphics[width = \columnwidth]{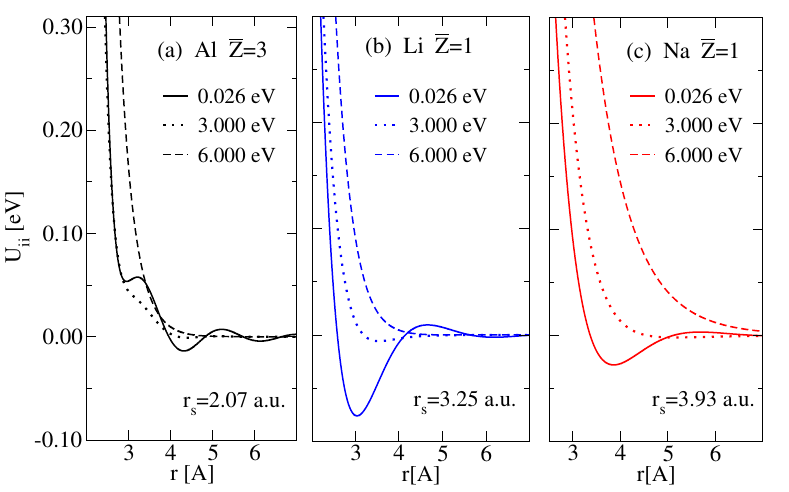}
\caption{(Color online) Two-temperature ion-ion pair potentials for
 electrons at three different
temperatures and ions at $T_i = 0.026$ eV (300 K), for 
(a) Al, (b) Li, and (c) Na. 
}
\label{fig:PairPot}
\end{figure}

Examples of NPA ion-ion pair potentials  at different temperatures are
presented in  Fig.~\ref{fig:PairPot}.  At equilibrium or sufficiently low
$T_e$,  all three pair potentials display Friedel oscillations as discussed in
section \ref{sec-npa}. Hence it requires many neighbor shells to compute the
total pairwise ion-ion interaction energy with sufficient precision. For Li and
Na, we used 8 shells  whereas  30 shells were necessary for the Al-Al interaction.  As  $T_e$ 
increases, the sharp Fermi surface breaks down, the discontinuity in $f(k)$ at
$k = k_F$ broadens, and oscillations disappear, yielding purely repulsive
Yukawa-screened potentials~\cite{YukawaScr}.

\subsection{$2T$ quasi-equilibrium phonon spectra.}
As the electrons get heated, the screening weakens and inter-ionic forces
become stronger; hence there is an interest in computing the phonon spectra
although in many cases the phonon oscillation times may be comparable to the
lifetime of the UFM system. 
Once the 2TPP is constructed for the desired $T_e$, the phonon spectra are
 easily calculated by the diagonalization of the dynamical matrix
 \cite{AshcroftBk}
\be
\mathbf{D}(\mathbf{k})=\sum_i\mathbf{D}(\mathbf{R}_i)e^{-i\mathbf{k}\cdot \mathbf{R}_i} \quad 
\ee
where the elements of the harmonic matrix $\mathbf{D}(\mathbf{R})$ are given by
\be
D_{\mu\nu}(\mathbf{R}) = \frac{1}{2}\sum_j\frac{\del^2 U_{ii}(\mathbf{R}_j)}{\del u_\mu(\mathbf{R})\del u_\nu(\mathbf{0})}
\ee
with $\mathbf{R}_j$ the position of the $j$th atom and $U_{ii}$ the
pair-potential of Eq.\ref{eq-pairpot}. From the $s$ eigenvalues
$\lambda_s(\mathbf{k})$ of $\mathbf{D}(\mathbf{k})$, the phonon frequencies are
given by $\omega_s(\mathbf{k})=\sqrt{\lambda_s(\mathbf{k})/M}$ with $M$ the
mass of the ion. 
The resulting phonons are compared with the results from ABINIT-DFT
simulations employing density-functional perturbation
theory~\cite{bib201,bib202} (DFPT), which determines the second derivative of
the energy using the first-order perturbation wavefunctions. We used the common
crystal structure for each metal, i.e.,  face-centered cubic (FCC) for Al and
body-centered cubic (BCC) for Li and Na, with their room temperature lattice
parameters $a = 4.05\ \AA, 3.49\ \AA$, and $4.23\ \AA$, respectively.

\begin{figure}[h]
\includegraphics[scale=1]{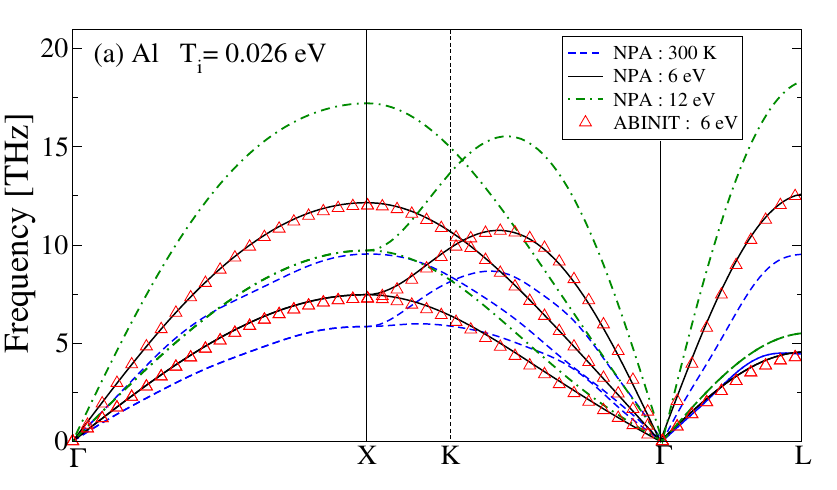}
\includegraphics[scale=1]{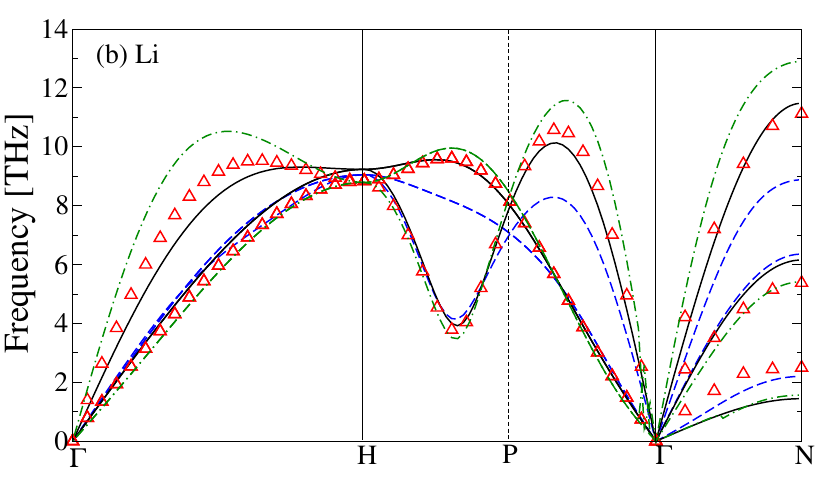}\vspace{0.1cm}
 \includegraphics[scale=1]{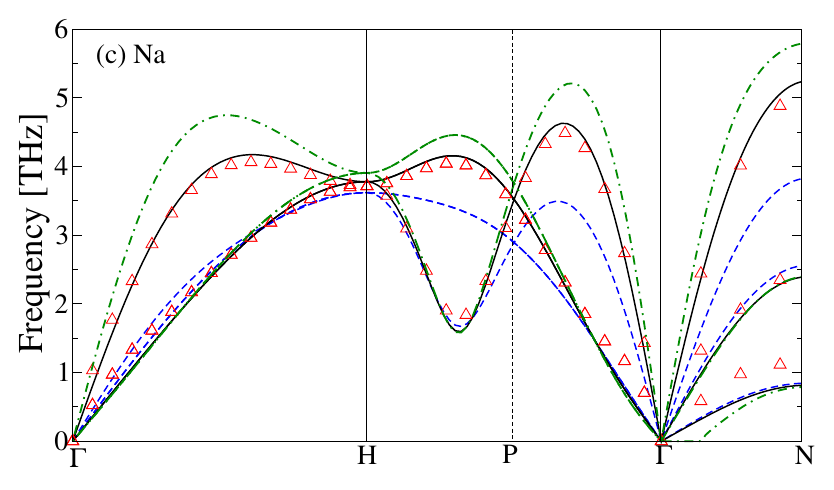}   
\caption{(Color online) Quasi-equilibrium phonon spectra at $T_e=6$ eV  obtained
with NPA and with ABINIT for (a) Al, (b) Li, and (c) Na. The NPA equilibrium
phonon spectra at 300 K are shown to illustrate the effect of increasing $T_e$
(dashed lines). }
\label{fig:NonEqPhonon}
\end{figure}

Quasi-equilibrium phonon dispersion relations at $T_e=6$ eV using the two 
methods are presented in Fig.~\ref{fig:NonEqPhonon}  with the NPA equilibrium
phonons as reference to illustrate important modifications in the spectra. In
addition,  NPA quasi-equilibrium phonon spectrum at $T_e = 12$ eV are also
presented by which temperature DFPT becomes prohibitive. The excellent accord
between the NPA and experimental equilibrium phonon spectra at low temperatures
has already been demonstrated and shows the meV accuracy of the NPA
calculations even at low temperatures~\cite{CPP-Harb}. This regime can be hard to model as noted by Blenski {\it et al.}~\cite{Blenski2013} when, for example, working on  
Al at normal density and at low $T$ within another model.

For the three systems in this study, the two methods (NPA and DFPT) predict very
similar $2T$ phonon spectra, thus reconfirming the  $2T$ NPA  calculations and
corroborating the DFPT calculations at finite $T$.  This is important as there
are as yet no experimental observations of UFM phonon spectra.  In the case
of Al, we observe a large increase in frequencies, as high as $32\%$ for
longitudinal (L) modes, which supports the ``phonon hardening'' theory. However,
we notice that transverse (T) branches in the $\Gamma-L$ region are barely
affected by the electron heating, as was also noted by Recoules~\cite{Recoules}.
In the case of Li and Na, we find that the spectral modifications are more
complex than the `homogeneous' increase found for Al; here, an important increase
in the L-branch in the middle of the $\Gamma-H$ region takes place, whereas there
is no change at the symmetry point $H$.  No modifications to T-branches are
noticed in this region. In the region $H-\Gamma$, the L-branch frequencies
increase in the middle of the region $H-P$ but remain unchanged at the symmetry
point $H$. For the T-branch, an increase is noticeable at the maximum in the
region $P-\Gamma$ whereas no change affects the minimum in the region $H-P$. In
the region $\Gamma-N$ and for the L-branch, we observe  the overall largest
increase of $29\%$ and $37\%$ for Li and Na, respectively, whereas frequencies of
T-modes are only slightly modified.

\subsection{$2T$-quasi-equilibrium equation of state.} 
A system in its initial equilibrium configuration ($T_i=T_e=T_r$) rapidly reaches
a new UFM state with $T_i$ remaining near $T_r$ while $T_e$ increases. 
However, since the ion motion within the time of arrival of the probe pulse is
negligible, the pressure builds up essentially isochorically due to electron
heating. 
 
\begin{figure}[h]
\includegraphics[scale=1]{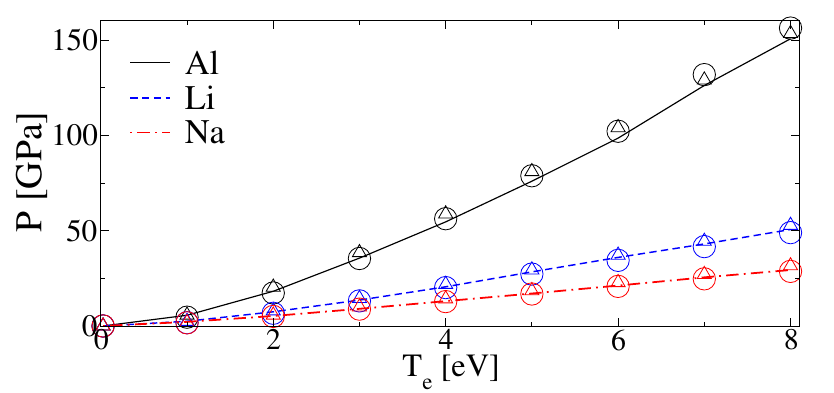}
\caption{(Color online) Quasi-equilibrium pressures  obtained with the NPA
(lines), ABINIT (circles), and VASP (triangles)  for Al, Li, and Na.}
\label{fig:Pressure}
\end{figure}

In  Fig \ref{fig:Pressure}, we compare the pressure calculated with the NPA model
with ABINIT and VASP simulations. In the latter, we used an energy cut-off of
1630 eV for the plane-wave basis, with 60 energy bands to capture finite-$T$
effects. In  ABINIT simulations, we used  norm-conserving (NC) pseudopotentials
with the $T=0$ Perdew-Burke-Ernzerhof (PBE) XCF within the generalized gradient
approximation (GGA). In VASP, we employed  projected-augmented-wave (PAW)
pseudopotentials with the PBE XCF for Li and Na, and the Perdew-Wang (PW) $T=0$
XCF for Al. With both codes, pseudopotentials were chosen specifically to
simulate $\bar{Z}$=3 valence electrons for Al, and $\bar{Z}$=1 for Li and Na as
the core electrons remain bound, and at the ion temperature. This is an important
aspect discussed in subsection~\ref{pseudo.sec}.

We find that, for all three metals, calculations using NPA, ABINIT and VASP 
predict nearly identical pressures with small deviations only at high $T_e$. At
$T_e=8$ eV, the maximum difference between all model is 9 GPa, 4 GPa and 3 GPa
for Al, Li,  and Na, respectively. Thus, the results from the extension of the
NPA model to the $2T$ regime confirms the usability of the  solid-state codes at
least up to 6 eV on the one hand, and on the the other hand the validity of the
NPA approach. However,  since NPA uses a finite-$T$ XC-functional whereas
\textit{ab initio} simulations do not, the effect of such finite-$T$ corrections 
will be reviewed  in section \ref{discussion}.

The computational efficiency and accuracy of the NPA
approach make it a valuable tool for studying WDM and other
complex systems where iterative computations of materials properties like
$2T$ EOS, $2T$ specific heat, transport properties, opacities,
energy-relaxation times, etc., are needed as the system evolves with
time, since mean ionization, pair-potentials and structure factors are
readily obtained. A few minutes on a desktop computer is
sufficient in NPA calculations to generate accurate results
which require long and intensive computations with DFT+MD.

\section{Discussion.}
\label{discussion}
\subsection{Crystal-lattice stability.}
\begin{figure}[h]
\includegraphics[scale=1]{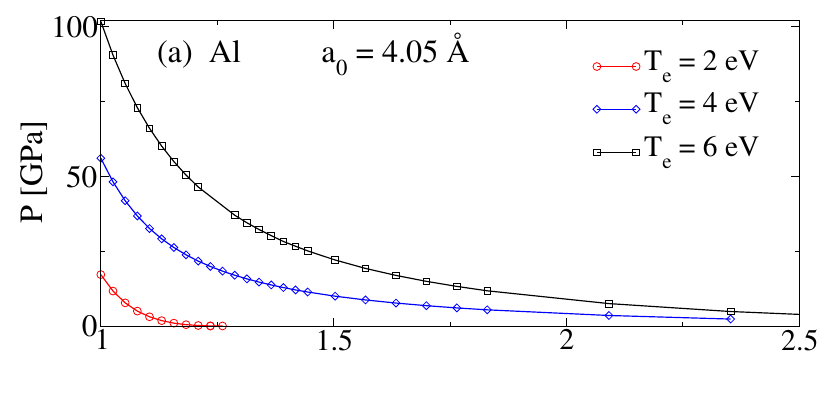}\vspace{-0.4cm}
\includegraphics[scale=1]{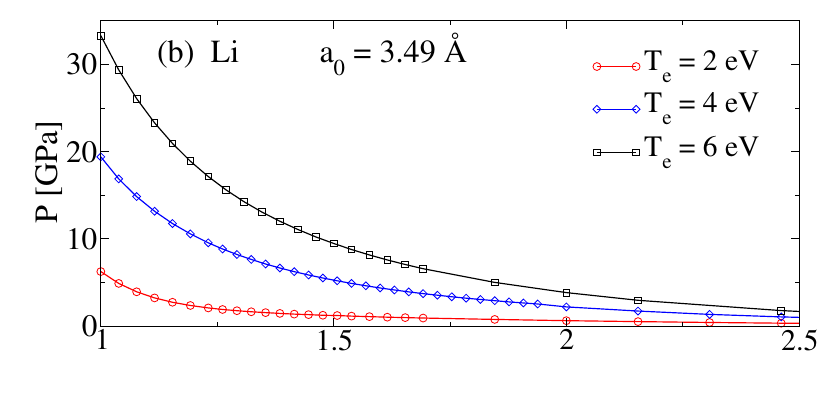}\vspace{-0.4cm}
\includegraphics[scale=1]{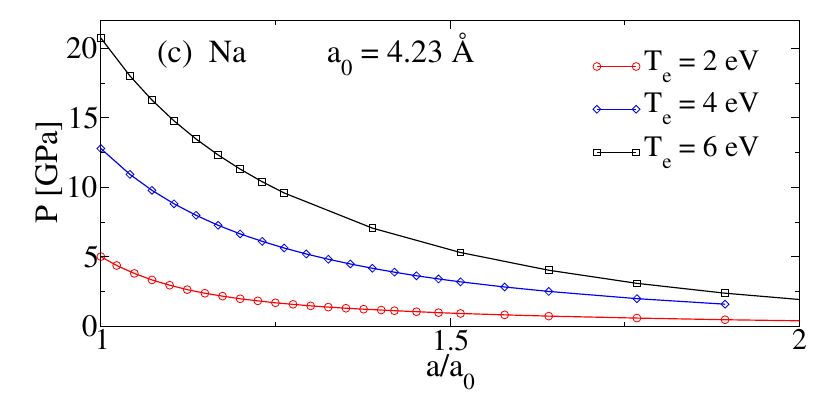}   
\caption{(Color online) Total pressure of the solids as a function of the lattice
parameter of the crystal relative to the room-temperature value $a_0$   for (a)
Al, (b) Li, and (c) Na.  }
\label{fig:LatticeExp}
\end{figure}

As electrons absorb the laser energy (within fs timescales) and heat up to $T_e$,
the internal pressure of the system becomes very high as discussed in section
III-C. In metals, the thermal expansion is also caused by the  free-electron
pressure. We studied the crystal stability of the solids as a function of lattice
expansion; the results are presented in Fig.~\ref{fig:LatticeExp}.

For Al at $T_e=2$ eV, we find that a moderate expansion $a/a_0 = 1.24$ is
sufficient to reduce the pressure back to zero, indicating that the crystal may
appear stable if the timescale needed for such lattice motion is available before
the UFM breaks down. However, in all other cases,  the pressure goes to zero
only asymptotically with increasing lattice parameter, suggesting that such 
UFM crystals are unstable. Such thermal expansions or spontaneous
fluctuations  lead to the `explosive' breakdown of the solid on ps timescales.
However, since UFM conditions are reached in fs timescales, the ions remain
essentially in their initial positions and (as already noted) no net linear
forces act upon them due to crystal symmetry. They remain trapped in a stronger
harmonic potential leading to hardening of most of the phonon branches. The
physical reason for the hardening at  increased $T_e$ is the decreased screening
of ion-ion interactions by the hotter electron gas.
\subsection{``Phonons'' and surface ablation.}
The UFM system is under very large pressure and the ion-ion $2T$PP is purely
repulsive unless $T_e$ is small (cf.\ Fig.~\ref{fig:PairPot}). The discussion in
terms of phonons may become inapplicable at higher $T_e$ due to non-zero ablation
forces acting on ions in typical UFM samples (0.1-1$\mu m$ thick). An ideal
periodic lattice implies that the linear derivative of the total potential is
zero because the crystal is isochorically constrained by the  external pressure.
The phonons of UFM ``exist'' only within this artifice. Small thermal
`Debye-Waller' type ionic displacements $u$ (with a mean value $\langle u\rangle
= 0.2 \AA $ at 300K for Al, retained in the UFM) do not render the periodic
UFM unstable, and  slightly split the degeneracy of transverse branches.

However, pump-probe experiments use very thin metal films. Crystal symmetry is
broken and large uncompensated forces act  at the  surface of the films; as a
result, the surface layer and successive layers ablate. We calculated the
ablation force $F_\text{abl}^\text{VASP}$ on an FCC-(100) Al surface and the two
inner layers using the VASP code with the Al surface reconstructed as happens for
the cold  surface at 0K.  Five layers of Al and 5 layers of vacuum were used for
evaluating the Hellman-Feynman forces on the surface atoms. The NPA method is
beyond its regime of validity  since the charge density at a surface is not
uniform. However, the NPA pressure is the force per unit area at the bounding
(100) surface, with one ion per unit area. This is used as the NPA estimate of
the ablation force $F_\text{abl}^\text{NPA}$. The forces on the inner neighbor 
and next-neighbor layers calculated from VASP at $T_e=$6 eV were 3\% and 0.02\%
respectively of the force on the surface layer. The surface force  $F_\text{abl}$
determines an approximate ``ablation time'' $\tau_\text{abl}$, the time needed
for the surface plane to move by an inter-plane distance  ($a/2$  in the case of
Al). This  $\tau_\text{abl}$  estimate makes some assumptions, e.g.,
$F_\text{abl}$ to be constant over  $a/2$, with no movement of  inner layers. To
verify if phonons can form within such timescales, we compare $\tau_\text{abl}$
with the shortest time  for an ion oscillation  $\tau_\omega$ at the highest
phonon frequency for the [100] direction; the results are presented in
Table~\ref{tabPhon}. 
\begin{table}[h]
\caption{The ``ablation force'' $F_\text{abl}$ 
 and the ``ablation  time''   $\tau_\text{abl}$ for the (100) surface
 of an  Al slab from VASP and NPA at three different electron temperatures
 $T_e$ and lattice temperature $T_i=0.026$ eV.
The fastest [100] phonon oscillation time $\tau_\omega$ is also given for
each $T_e$.}
\vspace{0.2cm}
\label{tabPhon}
\begin{tabularx}{\columnwidth}
{>{\centering}X>{\centering}X>{\centering}X>{\centering}X>{\centering}X>
{\centering}X}
\hline\hline
$T_e$ & $F_\text{abl}^\text{NPA}$ & $F_\text{abl}^\text{VASP}$ &
 $\tau_\text{abl}^\text{NPA}$& $\tau_\text{abl}^\text{VASP}$ & 
$\tau_\omega$ \tabularnewline
eV & eV/$\AA$ & eV/$\AA$ & fs & fs & fs \tabularnewline
\hline
2.00 & 0.91 & 0.90 & 111  & 111   & 105   \tabularnewline
4.00 & 2.75 & 2.70 & 63.9 & 64.2  &  92.6 \tabularnewline
6.00 & 5.03 & 4.70 & 47.1 & 48.6  &  80.6 \tabularnewline
\hline\hline
\end{tabularx}
\end{table}

As  $T_e$ increases, phonons ``harden'' and $F_\text{abl}$ increases.  In order
to observe the  ``hardening'' of phonons on any measurement, a probe time
$\tau_\text{pr}$ such that $\tau_\omega  < \tau_\text{pr} < \tau_\text{abl}$ is
required. However, for sufficiently high $T_e$  (e.g., above $\sim 2$ eV for Al),
the $F_\text{abl}$ are strong enough to make $\tau_\text{abl} < \tau_\omega$.
Hence the ion oscillations  have no time to build up and it is probably
impossible to satisfy the time constraint enabling the observation of hardened
phonons. The phonon  concept  itself becomes  misleading for thin UFM
films.  Interpreting experiments when $\tau_\text{pr} >\tau_\text{abl}$ may
require explicit inclusion of  surface ablation corrections in the theory used 
for analyzing optical data (e.g., in the Helmholtz equations).
\subsection{Finite-$T$ exchange and correlation.}
\begin{figure}[!h]
\includegraphics[width=0.95\columnwidth]{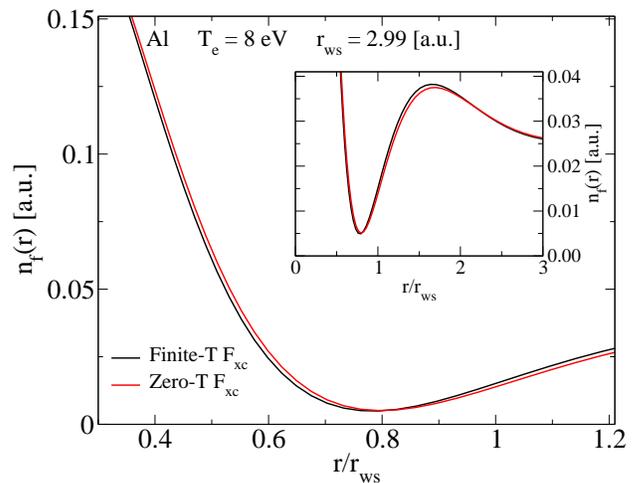}
\caption{(Color online) The NPA free-electron density $n_f(r)$ for
Al$^{3+}$ at density $\rho=2.7$ g/cm$^3$, with $T_e = 8$ eV and $T_i =
0.026$ eV,  calculated using XC at finite-$T$ and at $T=0$. 
The inset shows the density
for larger $r/r_\text{ws}$. 
}
\label{fig:Thompson}
\end{figure}

In the NPA model, we used the finite-$T$ XCF of PDW and assessed the importance
of such corrections in the temperature regime studied here. The valence  density,
or ``free''-electron density $n_f(r)$ of the solid at $T_e>T_i$ is the key
quantity for the NPA model.  In Fig.~\ref{fig:Thompson}, we present the $n_f(r)$
obtained using the PDW finite-$T$  XCF with that obtained from the zero-$T$ XCF.
Even though the correction is small, it may be of importance in some
circumstances, e.g., x-ray Thomson scattering spectra, and hence there is no
reason to neglect it. The difference between the $T=0$ XCF and the finite-$T$ XCF
increases with $\theta=T/E_F$ at first, and it rapidly and asymptotically goes to
zero as $\theta>1$ and as $T\to\infty$. Hence the more important consequences of
using finite-$T$ XCF should occur in the partially degenerate regime $0<\theta <
1$.  

\begin{figure}[!h]
\includegraphics[width=0.95\columnwidth]{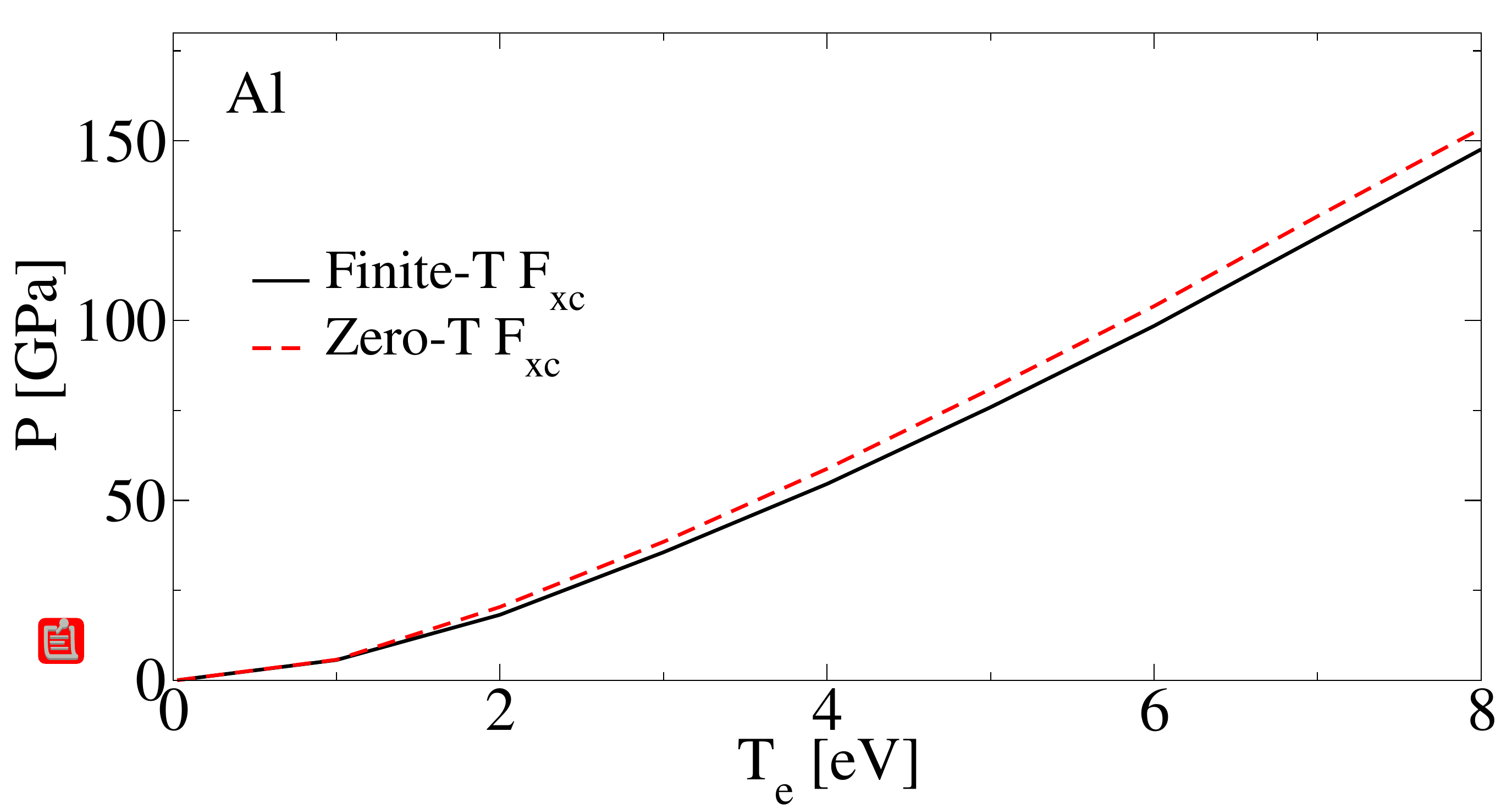}
\caption{(Color online) Comparison between the pressure of Al in the UFM regime
computed via the NPA model with the finite-$T$ $F_{xc}$ and with the zero-$T$
$F_{xc}$.}
\label{fig:Al_P_Exc}
\end{figure}

The finite-$T$ XCF is present in two  contributions to the pressure, namely the
electron-electron interacting linear response function $\chi(k,T_e)$, which is
used to construct the pseudopotential and the pair potential, and the HEG
electron kinetic pressure. Although the finite-$T$ XCF has noticeable effects on
the pair potentials or on the energy spectrum of bound states, we observe that
overall thermodynamic effects are only slightly sensitive to such finite-$T$
corrections as can be seen in Fig.~\ref{fig:Al_P_Exc}. In fact, at $T_e = 8$ eV,
the finite$-T$ XCF only decreases the pressure in Al by $4\%$. Since individual
finite-$T$ contributions are considerable, this insensitivity to XCF comes from
the interplay of several terms. For instance, the electron pressure by itself
differs by about 10$\%$ in the regime $\theta \sim 0.8$,  but the overall
pressures obtained from $T=0$ and finite-$T$  NPA calculations differ by less
than $4\%$.

\subsection{Pseudopotential and mean ionization.}
\label{pseudo.sec}
Here, we discuss the importance of choosing the proper pseudopotential for
\textit{ab initio} simulations of UFM systems in the $2T$ model. 
The pump-laser frequency is normally chosen such that core electrons are not
excited  and remain strongly bound to the `cold' nuclei at temperature $T_i$.
Thus, only the $\bar{Z}$ valence electrons on each ion are heated to $T_e$ during
the irradiation. In DFT calculations, the electron temperature is used in a Fermi-Dirac distribution
for the occupation numbers of all electrons in the simulation. Thus,  if the chosen
pseudopotential includes more electrons than the typical number of valence
electrons, these core electrons will also be ``heated'' even if they should
not in order to simulate correctly UFM systems. Wrong predictions may result, e.g., for
the 2T pressure of the given UFM and its electronic specific heat.

To illustrate this point, we carried out ABINIT simulations using PAW
pseudopotentials which include $\bar{Z}$=3 and 9 valence electrons for Li
and Na, respectively. We also did NPA-DFT
calculations with {\em all}  electrons at $T_e$. In the NPA model, the mean
ionization $\bar{Z}=Z_n-n_b$ can be computed  as in Ref.~\cite{PDW2}. 
The $\bar{Z}$ as a function of $T_e$ is not an integer in the NPA but represents
an average over different ionization states as discussed in Ref.~\cite{PDW95}.

In the case of Al and Li, the NPA predicts that $\bar{Z}$ is unaffected  for
$T_e< 8$ eV, relevant to UFMs. Pressure should also be unchanged, which is
exactly what we obtain with the ABINIT simulations of Li using the
all-electron PAW pseudopotential. However, in the case of Na,  $\bar{Z}$
starts to increase around $T_e=3$ eV up to $\bar{Z}=1.49$ at $T_e=8$ eV  (see
Table.~\ref{zbar-na.tab}). The increase in $\bar{Z}$ is accompanied by a
decrease in the occupation of the 2$p$ level as electrons are promoted to the
continuum. The decreased screening in the core (both due to increase of $T$
and due to the decrease in the number of core electrons) leads to a {\it
decrease} in the radius of the $n$=2 shell. Hence, the increase of $\bar{Z}$
and the modification of the core levels do not lead to any ambiguity in
specifying $\bar{Z}$. %
\begin{table}[]
\caption{Mean ionization $\bar{Z}$, the 2$p$ Fermi factor,  and the 2$p$ mean
radius (a.u) for sodium  (normal solid density) are  given as a function of the
temperature $T$ in eV. The WS radius $r_{ws}$=3.3912 a.u. and hence
the core is compactly contained inside the WS sphere of Na for all values of $T$
investigated here.}
\vspace{0.2cm}
\label{zbar-na.tab}
\begin{tabularx}{\columnwidth}
{>{\centering}X>{\centering}X>{\centering}X>{\centering}X}
\hline\hline
$T$ & $\bar{Z}$ & $f_{2p}$ & $<r_{2p}>$  \tabularnewline
\hline
1.00 & 1.001 & 1.000  & 0.808  \tabularnewline
3.00 & 1.004 & 0.999  & 0.804  \tabularnewline
5.00 & 1.104 & 0.983  & 0.792  \tabularnewline
8.00 & 1.494 & 0.919  & 0.762  \tabularnewline
10.0 & 1.786 & 0.872  & 0.744  \tabularnewline
\hline\hline
\end{tabularx}
\end{table}

\begin{figure}[b]
\includegraphics[scale=1]{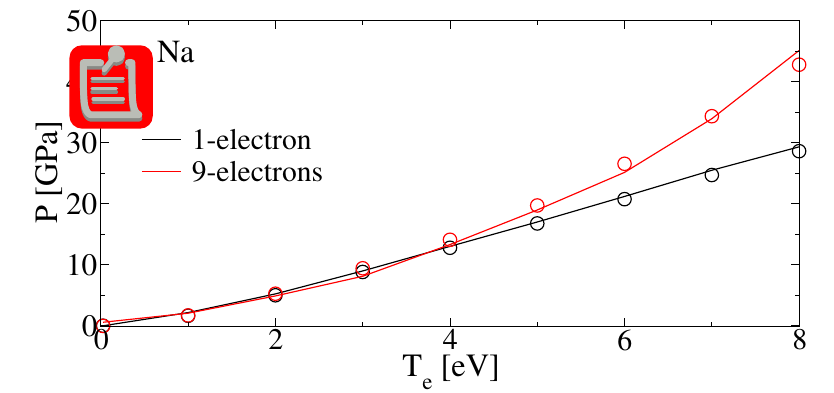}
\caption{(Color online)  Comparison between the pressure 
computed with the NPA (line) and with ABINIT (circles) when 
heating is applied to the valence electron of Na only or to all electrons  
(9 electrons in the ABINIT simulations).}
\label{fig:NaPressurePAW}
\end{figure}

We computed the pressure
with the NPA model including the changed $\bar{Z}$ and compared it with the
ABINIT simulations of Na using the nine-electron PAW pseudopotential. Results
are  presented in Fig.~\ref{fig:NaPressurePAW}.
We find that, at $T_e=8$ eV, the pressure, when heating of some core electrons
is included, is  $54\%$ higher than the correctly calculated value.  The use
of `all-electron' codes for the study of UFM in the $2T$ state suffers from
this pitfall of not selecting the physically appropriate $\bar{Z}$ and the
corresponding pseudopotential.  When suitable pseudopotentials are not
available for DFT+MD calculations, one possibility is to use only the relevant
part of the electron density of states (DOS) that is assigned to the free
electrons  on the basis of $\bar{Z}$, when pressure and related properties are
computed.  For instance, when calculating the specific heat of `free
electrons' for use in UFM studies, the  `free-electron' DOS used in the
calculations  should be consistent with the number of actual free electrons
that couple with the laser. In a metal like gold (not studied here), even
though a pseudopotential with 11 valence electrons is needed, the DOS used for
evaluating the  electron specific  heat for $T_e<2$ should be only for
$\bar{Z}=1$. The optical properties of gold (see ref.~\cite{ChristyJon}) show
that the $d$-shell  couples to light only  when the interband  threshold
energy ($\sim$ 2 eV) is exceeded. In the case of gold, the 5$d$ shell
hybridizes with the continuum electrons (nominally made up of 6$s$ electrons)
and extends outside the Au-Wigner-Seitz sphere until the $s-d$ transition
threshold  ($\sim $2 eV) is reached. Hence, at low temperatures the NPA model 
with its `one-center' formulation cannot be used for gold at normal density.
Similarly, WDM systems with bound states extending outside the Wigner-Seitz
 sphere cannot be treated
unless explicit multi-center electron-ion correlation terms are included.

\subsection{Local pseudopotential for Li.}

The Li pseudopotential used in the NPA is a {\it local} pseudopotential,
whereas it is widely found in the context of large DFT codes that Li almost
always needs a non-local pseudopotential. Even in early studies of phonons,
a nonlocal pseudopotential was used by Dagens, Rasolt, and Taylor~\cite{DRT},
and yet the Li phonons at room temperature they obtained were less
satisfactory than for, say, sodium.  We have already shown that the NPA
pair potential based on a local pseudopotential quite adequately
 reproduces the Li phonons at
room temperature and high temperature at normal density,
but not as accurately as for aluminum or sodium.  Hence it is of interest to
test the robustness of the Li pseudopotential and pair potential  at
higher compression  by calculating the Li-Li $g(r)$ using the NPA
potentials.  Here we use the MHNC method 
where a bridge term is included using the Lado-Foiles-Ashcroft (LFA)
criterion~\cite{LFA-ChenLai92} which is based on the Gibbs-Bogoliubov
inequality for the free energy of the system. The MHNC assumes radial symmetry and lim-
its us to ``simple-liquid''
structures.
\begin{figure}[h]
\includegraphics[width= 0.95\columnwidth]{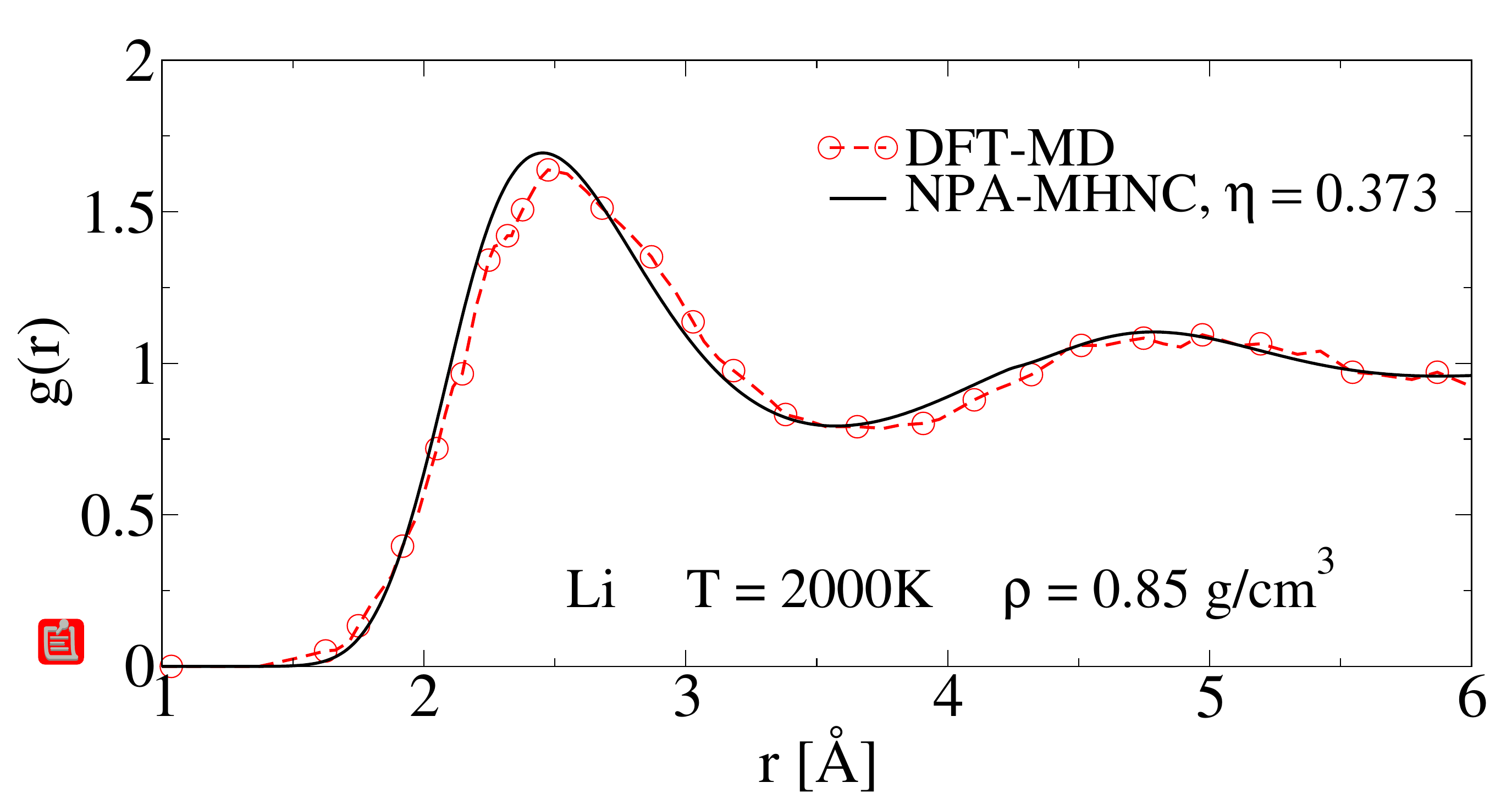}
\caption{(Color online) The Li-Li NPA-MHNC pair distribution function
$g(r)$ at 2000K (0.173 eV), $\rho=$ 0.85 g/cm$^3$, compared
with the  $g(r)$ of Ref.~\cite{kietzmann2008}.
}
\label{Li-gr-p85g.fig}
\end{figure}

Since Li becomes a complex liquid with clustering effects at high
compressions~\cite{bonev2008}, we consider a compression of $\sim 1.6$
and compute the PDF for Li at 0.85 g/cm$^2$ and at 2000K (0.173 eV) for which
 results are available from Kietzmann {\it et
al.}~\cite{kietzmann2008}. The LFA criterion yields a hard-sphere packing
fraction $\eta=0.371$ to model the bridge function. The resulting NPA-MHNC
$g(r)$ is displayed together with the  $g(r)$ of Ref.\ ~\cite{kietzmann2008}
 in Fig.~\ref{Li-gr-p85g.fig}.
 We find
that the simple but state-dependent {\it local} pseudopotential
constructed from the free-electron charge pileup at a Li nucleus is
adequate to calculate phonons (i.e, requiring an accuracy of meV energies),
as well as the Li-Li PDFs up to moderate compressions and
 high coupling constants $\Gamma$. 
\subsection{Comparison between equilibrium WDM and UFM EOS.}
In UFM, the internal pressure mainly results from the hot
electron subsystem since ions remain close to their initial temperature
$T_r$. Here, we investigate the difference in the pressure between the
quasi-equilibrium UFM regime ($T_i\neq T_e$) and the equilibrium WDM regime which
will usually be in a liquid or plasma state  with $T_i=T_e$. In DFT codes it is possible
to simulate liquids  by computing forces among ions and the MD evolution of
the positions of the $N$ ions in the simulation cell.
 However, to obtain
reasonable statistics, one needs to use  a supercell containing as many ions
as possible, thus reducing considerably the first Brillouin zone and increasing
the required number of electronic bands to be included. As mentioned
earlier, the number of bands needs to be even larger in order to
simulate $T_e$ via a Fermi-Dirac distribution. As examples, to obtain
reasonably good band occupations for a system of 108 Al atoms at room
density, 360 bands at $T_e=1$ eV are required, and this number grows to 1200 
at $T_e= 5$ eV. Thus, since computing repeatedly at every MD step
 a high number of bands in DFT codes is
computationally very demanding, it becomes
prohibitive at higher temperatures. Such a problem does not occur
in the NPA model as only one DFT calculation at a single nucleus
is required to construct the ion-ion pair potential. The structure factors
may be computed using MD, or with MHNC equation for 
simple liquids. 

\begin{figure}[h]
\includegraphics[width= 0.95\columnwidth]{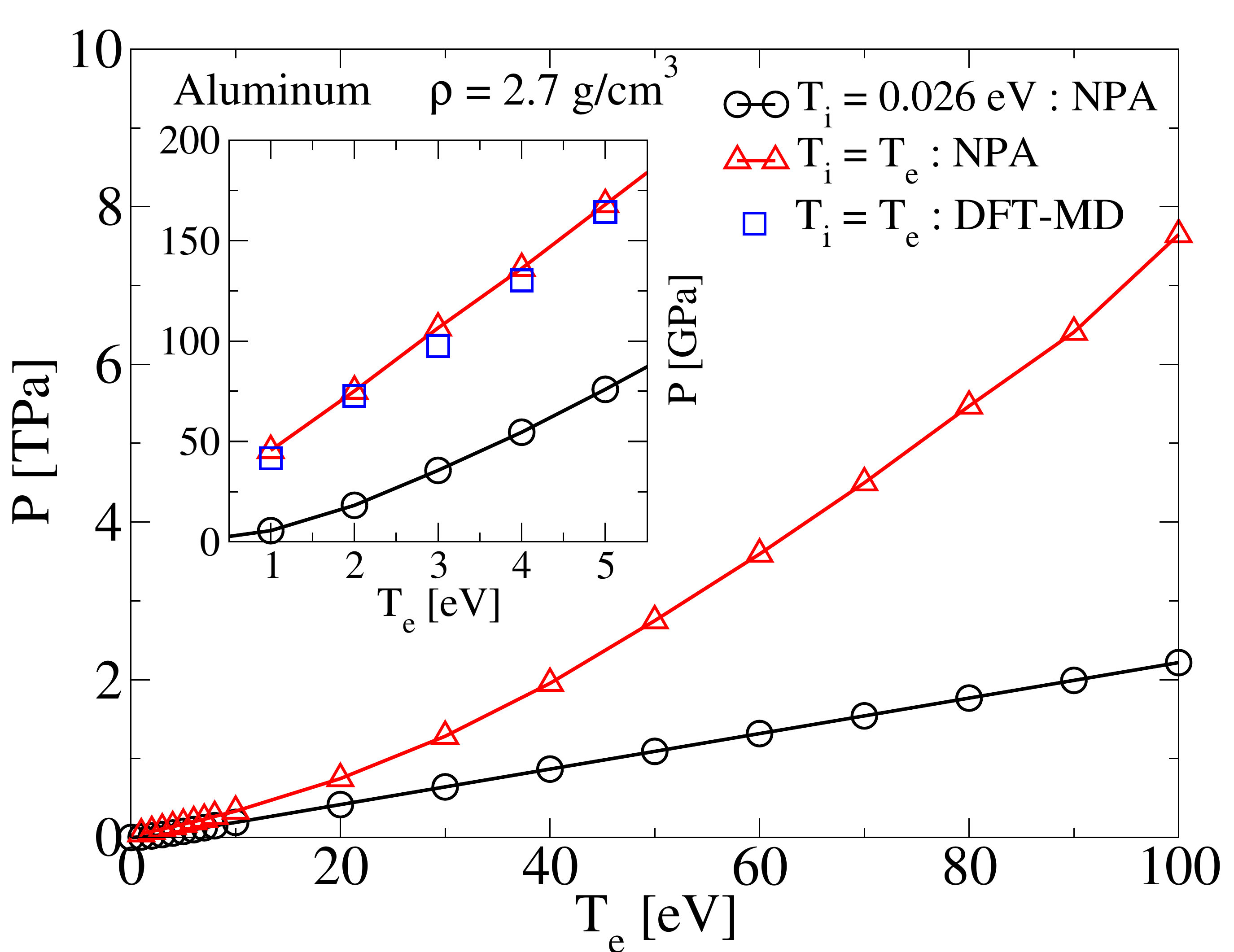}
\caption{(Color online) Comparison of the NPA isochoric pressures 
 for the UFM system and the equilibrium liquid system.
Inset : Comparison of the NPA  pressures in the low-$T$ regime
 where DFT+MD is practical.
}
\label{Al_Liq_P.fig}
\end{figure}

The comparison of the pressure from UFM and equilibrium WDM
 is presented 
in Fig.~\ \ref{Al_Liq_P.fig}. The equilibrium WDM  pressure is
 much higher than the UFM value. Furthermore,
 the DFT-NPA calculation is in agreement with NPA up to $T_e = 5$ eV 
(the limit of our DFT+MD simulation). This mutually
reconfirms the validity of the NPA as well as DFT+MD approaches
in the WDM regime. 

Since $\bar{Z}$ reaches $\sim$ 7 at $T_e\sim$100 eV,  codes for
simulating Al should employ pseudopotentials that include more
electrons than the 3 valence electrons valid at low temperatures.
 Simulations with high $\bar{Z}$ values will greatly increase the
computational load and such calculations become prohibitive. Hence NPA
methods or orbital-free Hohenberg-Kohn
 methods become relevant~\cite{karaisev2014}. The latter
do not however provide energy spectra and details of the bound electrons.

\section{Conclusion.}

In order to describe physical properties of UFM, we
examined applications of the NPA model within  the
two-temperature quasi-equilibrium model. We computed phonons, as well as
the pressure resulting from the heating of free electrons. The
excellent accord between such NPA calculations and DFT simulations using the
ABINIT and VASP codes reconfirms the use of the NPA in this  regime. As the
internal pressure increases due to the heating of electrons by the ultrafast
laser pulses, we explicitly showed that the phonon picture does not have much
physical meaning, especially for thin WDM samples, even if frequencies could
be computed using the harmonic approximation.  As the NPA approach has
negligible computational cost compared to standard DFT codes, it is a valuable
tool for swiftly and accurately calculating important WDM properties such as mean
ionization, pair potentials, structure factors, phonons, x-ray Thomson
scattering spectra, electron-ion energy relaxation, conductivity, etc..

\section*{Acknowledgments.} 
This work was supported by grants from the
Natural Sciences and Engineering Research Council of Canada (NSERC)
and the Fonds de Recherche du Qu\'{e}bec - Nature et Technologies (FRQ-NT).
We are indebted to Calcul Qu\'{e}bec and Calcul Canada for generous
allocations of computer resources.

\end{document}